\pdfoutput=1
\documentclass[authoryear]{article}

\usepackage{pdflscape}
\usepackage{ltablex}
\usepackage{geometry}
\usepackage{lineno}
\usepackage{hyperref}
\usepackage{tabularx}
\usepackage{multirow}
\usepackage{booktabs}
\usepackage{graphicx}
\usepackage{ccaption} % allow to continue caption on next page
\usepackage[colorinlistoftodos]{todonotes} % Allow comments to be visible on the pdf
\usepackage{floatpag} % Take out page numbering from appendix
\usepackage{longtable} % Span tables over multiple pages
\usepackage{pdfpages} % Add pdfs
\usepackage{rotating}
\usepackage{changepage} % allow margins for specific page to be changed
\usepackage{caption} % Allow captionoff
\usepackage[affil-it]{authblk} % LIbrary for affiliation
\usepackage[round]{natbib}
\usepackage{float} %force floats to stay in specific positions
\usepackage{fancyhdr}
\fancyheadoffset{0pt}
\def\keywordname{{\bfseries \emph Keywords}}%
\def\keywords#1{\par\addvspace\medskipamount{\rightskip=0pt plus1cm
\def\and{\ifhmode\unskip\nobreak\fi\ $\cdot$
}\noindent\keywordname\enspace\ignorespaces#1\par}}

\begin{document}
\title{\vspace{-2cm}\textbf{Analysis of an Automated Machine Learning Approach in Brain Predictive Modelling: A data-driven approach to Predict Brain Age from Cortical Anatomical Measures}}

\author[1]{Jessica Dafflon\thanks{ \texttt{jessica.dafflon@kcl.ac.uk}}}
\author[2,3]{Walter H. L. Pinaya}
\author[1]{Federico Turkheimer}
\author[1]{James H. Cole}
\author[1]{Robert Leech}
\author[4]{Mathew A. Harris}
\author[5,6]{Simon R. Cox}
\author[4]{Heather C. Whalley}
\author[4]{Andrew M. McIntosh}
\author[1]{Peter J. Hellyer\thanks{\texttt{peter.hellyer@kcl.ac.uk}}}

\affil[1]{Department of Neuroimaging, Institute of Psychiatry, Psychology and Neuroscience, \newline King's College London, UK}
\affil[2]{Department of Psychosis Studies, Institute of Psychiatry, Psychology and Neuroscience, \newline King's College London, UK}
\affil[3]{Center of Mathematics, Computation and Cognition, Universidade Federal do ABC, Brazil}
\affil[4]{Division of Psychiatry, University of Edinburgh, UK}
\affil[5]{Lothian Birth Cohorts group, Department of Psychology, University of Edinburgh, UK}
\affil[6]{Scottish Imaging Network, A Platform for Scientific Excellence (SINAPSE) Collaboration, Edinburgh, UK}

\renewcommand\Authands{ and }
% Change size of the affiliation font
\renewcommand\Affilfont{\itshape\small}

\maketitle

%%%%%%%%%%%%%%%%%%%%%%%%%%%%%%%%%%%%%%%%%%%%%%%%%%%%%%%%%%%%%%%%%%%%%
\vspace{-1.5cm}
%%%%%%%%%%%%%%%%%%%%%%%%%%%%%%%%%%%%%%%%%%%%%%%%%%%%%%%%%%%%%%%%%%%%%%%%%%%%%%%%%%%
\begin{abstract}
The use of machine learning (ML) algorithms has significantly increased in
    neuroscience. However, from the vast extent of possible ML algorithms, which
    one is the optimal model to predict the target variable? What are the
    hyperparameters for such a model? Given the plethora of possible answers to
    these questions, in the last years, automated machine learning (autoML) has
    been gaining attention. Here, we apply an autoML library called TPOT which
    uses a tree-based representation of machine learning pipelines and conducts
    a genetic-programming based approach to find the model and its
    hyperparameters that more closely predicts the subject's true age. To
    explore autoML and evaluate its efficacy within neuroimaging datasets, we
    chose a problem that has been the focus of previous extensive study: brain
    age prediction. Without any prior knowledge, TPOT was able to scan through
    the model space and create pipelines that outperformed the state-of-the-art
    accuracy for Freesurfer-based models using only thickness and volume
    information for anatomical structure. In particular, we compared the
    performance of TPOT (mean accuracy error (MAE): $4.612 \pm .124$ years) and
    a Relevance Vector Regression (MAE $5.474 \pm .140$ years). TPOT also
    suggested interesting combinations of models that do not match the current
    most used models for brain prediction but generalise well to unseen data.
    AutoML showed promising results as a data-driven approach to find optimal
    models for neuroimaging applications.
\end{abstract}

%%%%%%%%%%%%%%%%%%%%%%%%%%%%%%%%%%%%%%%%%%%%%%%%%%%%%%%%%%%%%%%%%%%%%
\keywords{
\texttt{predictive modelling} \and automated machine learning \and age prediction \and neuroimaging}
%%%%%%%%%%%%%%%%%%%%%%%%%%%%%%%%%%%%%%%%%%%%%%%%%%%%%%%%%%%%%%%%%%%%%

%%%%%%%%%%%%%%%%%%%%%%%%%%%%%%%%%%%%%%%%%%%%%%%%%%%%%%%%%%%%%%%%%%%%%
% Add content
%%%%%%%%%%%%%%%%%%%%%%%%%%%%%%%%%%%%%%%%%%%%%%%%%%%%%%%%%%%%%%%%%%%%%
%%%%%%%%%%%%%%%%%%%%%%%%%%%%%%%%%%%%%%%%%%%%%%%%%%%%%%%%%%%%%%%%%%%%%%%%%%%%%%%%%%%
% Introduction
%%%%%%%%%%%%%%%%%%%%%%%%%%%%%%%%%%%%%%%%%%%%%%%%%%%%%%%%%%%%%%%%%%%%%%%%%%%%%%%%%%%

\section{Introduction}

The last few decades have seen significant progress in neuroimaging
methodologies and techniques focused on identifying structural and functional
features of the brain associated with behaviour. These methods, have been widely
applied to assess differences at a group level between, for example, clinical
groups. However, group-level statistics are limited and fail to make inferences
that are applicable to the individual. With the advance of machine learning (ML)
algorithms and their increased application in neuroimaging, the field is rapidly
becoming more focused on exploring relationships between individual difference
and behaviour, as well as, developing clinically relevant biomarkers of
disorders \citep{Pereira-2009Machine, Liem-2017Predicting, Glaser-2019The,
Yarkoni-2017Choosing, Bzdok-2019Exploration}.

This recent shift was mainly due to the use of predictive modelling approaches,
consisting of using ML algorithms to learn patterns from features in a dataset
and to build an accurate model to predict an independent variable of interest in
unseen data. However, choosing a model  which is unsuitable for the statistical
distribution the underlying data leads to significant problems with
\emph{over}-estimation of the model and loss of generalisation. Secondly, the
sheer \emph{mass} of learning approaches that are available with a vast array of
different properties provides a bewildering set of choices for the practitioner;
each with advantages and disadvantages both in terms of generalisation and
computational complexity. This issue results in the occurrence of both type I
and II errors, simply as a result of picking an inappropriate analysis technique
for the underlying data. This is particularly problematic as new fields adopt
machine learning approaches, and the choice of the methodology is often based on
applications in other fields where data may have quite different statistical
properties - or indeed simply be the product of whichever technique is currently
in the zeitgeist.

The \emph{no free lunch principle} \citep{Wolpert-1997No} applied to ML,
suggests that there are no single estimator and parameter combinations that will
always perform well on every dataset. The selection of preprocessing steps, the
choice of the algorithm, the selection of features and the model's
hyperparameters are crucial and will vary with the task and data. Hence, the
optimal application of ML technology requires the answer to at least three
questions: What are the necessary preprocessing steps that should be performed
to prepare the data? Is there a way of reducing the feature space to only the
relevant features? Among the many available ML algorithms which one is the most
appropriate for the data under analysis? That these choices are often arbitrary
and defined only on \emph{prior}-wisdom, is a challenge for neuroimaging which
continues to face a significant replication crisis
\citep{Open-Science-Collaboration-2015PSYCHOLOGY.}.

ML algorithms vary greatly in both their properties, complexity and the
assumptions they make about the data they are applied to. They can be linear,
non-linear and optimise different functions to predict continuous (regression)
or categorical (classification) variables. Moreover, the performance of all ML
algorithms depends on the fine-tuning of its hyperparameters
\citep{Jordan-2015Machine}. In addition, feature extraction and feature
selection methods are often used in series to reduce or enhance data complexity
during the preprocessing stages of analysis. The consequence is that there are
potentially infinite combinations of approaches that can be taken to identify
relationships out of data. To cut through this complexity requires the
development of tools that can automatically select the appropriate (combination
of) preprocessing and ML techniques to apply to a dataset to highlight
relationships that are both generalisable and computationally efficient.

\begin{figure}
    \centering
    \includegraphics[width=\textwidth]{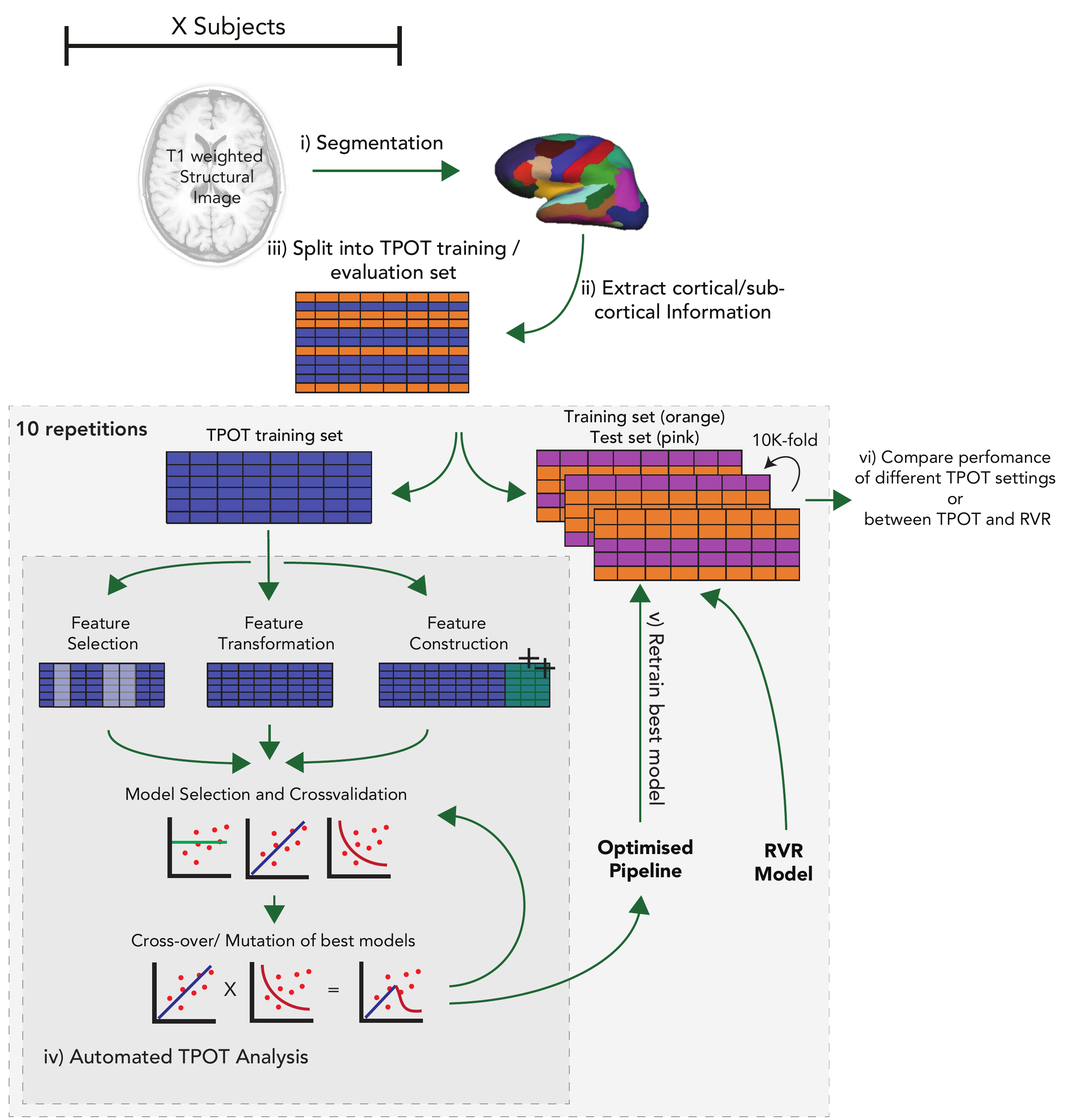}
    \caption{\textbf{Overview of experimental design:} The subject's structural
    MRI is used to create a parcellation of cortical and subcortical regions.
    The dataset was split into 2 independent sets: TPOT training set and evaluation
    set. The TPOT training set was passed to TPOT, which depending on the
    specified configuration performed feature selection, feature transformation,
    feature generation, or a combination of those and evaluated the model's
    performance. For each generation, a 10-fold cross-validation was performed
    and the best models for that specific generation were identified,
    crossed-over/mutated, and passed to the next generation. At the last
    generation, the pipeline with the lowest mean accuracy error was identified
    and returned by TPOT. We then retrained the optimised pipeline on the
    independent evaluation set and tested its performance using a 10-fold
    cross-validation. Finally, we compared the MAEs between different TPOT
    configurations and between TPOT and RVR.}
    \label{analysis_overview}
\end{figure}

In recent years, automated machine learning (autoML) has been gaining attention.
The aim of autoML is to take advantage of complexity in the underlying dataset
to help guide and identify the most appropriate model (and their associated
hyperparameters), optimising performance, whilst simultaneously attempting to
maximise the reliability of resulting predictions.  In this context, many
different autoML libraries have been developed.
Auto-WEKA~\citep{Thornton-2013Auto-WEKA:},
Auto-Sklearn~\citep{Feurer-2015Efficient} and Tree-based Pipeline Optimisation
Tool (TPOT)~\citep{Olson-2016Evaluation} are just a few examples.  While the
first two implement a hierarchical Bayesian method, the latter uses a tree-based
genetic programming algorithm. Due to its user-friendly interface and the
pipeline flexibility offered by the optimisation of a tree-based approach
\citep{automl-book}, we have chosen to evaluate TPOT's performance on this
problem.  The main idea behind the tree-based genetic programming is to explore
different pipelines (i.e. combination of different operators that perform
features selection, feature generation and model analysis) for solving a
classification or regression problem. This is done through a multi-generation
approach, starting from a collection of \emph{random} models. Based on the
performance and reliability of predictions at each generation those with the
highest performance will be \emph{bred} (i.e. combined or crossed-over), whilst
random \emph{mutations} of these models are also introduced. Therefore
combinations of models that maximise both performance and have lower complexity
survive and the 'best' candidate pipeline yielded by TPOT will consist of a
combination of models and preprocessing methods that are best suited to the
relationship being probed. Figure~\ref{analysis_overview} presents a high-level
schematics of our approach.

In this paper, we explore the application of TPOT as an autoML approach to
structural neuroimaging data. As a test-case, we evaluated its efficacy to
predict chronological age using structural brain data. Ageing is one factor
inducing major variability in brain structure. Grey matter atrophy, increase in
the ventricle sizes, cortical thinning are a few examples of structures that
alter while we age \citep{Hogstrom-2013The, Cole-2017Predictinga}. As
age-related changes can be detected with structural magnetic resonance imaging
(MRI) different machine learning models, have been trained to learn the
relationship between age and brain structure \citep{Franke-2010Estimating,
Aycheh-2018Biological, Cole-2015Prediction, Liem-2017Predicting,
Valizadeh-2017Age, Madan-2018Predicting, Becker-2018Gaussian}. The main idea
behind brain age studies is to find discrepancies between the predicted and
chronological age, which might be used as biomarkers
\citep{Cole-2017Predictinga}. As brain-age prediction has been extensively
studied and its accuracy can be evaluated against the reported model accuracies
the existing brain-age corpus \citep{Aycheh-2018Biological,
Cole-2017Predictingb, Franke-2010Estimating, Valizadeh-2017Age}, we used this
problem to test the settings, validity and limitations of autoML for imaging
applications in using a regression approach. In this study, we demonstrate that:
(1) the model's performance is highly dependent on the initial model population
defined by the initial model pool passed as a configuration and the population
size; (2) there is no single analysis model that predicts age with the highest
performance from the underlying structural imaging data; (3) models suggested by
TPOT outperforms relevance vector regressor (RVR), a state-of-the-art model used
to predict brain age. Therefore, TPOT can be used as a data-driven approach to
learn patterns in the data, to automatically select the best hyperparameters and
models in a researcher unbiased fashion to avoid common pitfalls from ML
algorithms such as overfitting.

%%%%%%%%%%%%%%%%%%%%%%%%%%%%%%%%%%%%%%%%%%%%%%%%%%%%%%%%%%%%%%%%%%%%%%%%%%%%%%%%%%
% Methods
%%%%%%%%%%%%%%%%%%%%%%%%%%%%%%%%%%%%%%%%%%%%%%%%%%%%%%%%%%%%%%%%%%%%%%%%%%%%%%%%%%%
\section{Methods}

\subsection{Subjects and Datasets} In this analysis, T1-weighted MRI scans from
N=10,307 healthy subjects (age range 18-89 years, mean age = 59.40) were
obtained from 13 publicly available datasets where each dataset used one or more
scanners to acquire the data. A summary of the demographics and imaging
information can be found in Table~\ref{table:demography} (for more details about
the BANC dataset see \citep{Cole-2017Predictingb}) and for the UK Biobank
\citep{Sudlow-2015UK, Alfaro-Almagro-2018Image}
((\url{https://biobank.ctsu.ox.ac.uk/crystal/crystal/docs/brain_mri.pdf}). From
the original n=2001 subjects present on the BANC dataset, we only used 1227
subjects and excluded all subjects from the WUSL Cohort after performing
Freesurfer quality control checks.

\subsection{MRI Preprocessing} Using the recon-all pipeline in Freesurver
version v6.0~\citep{Dale-1999Cortical}, individual T1-weighted MRI images were
preprocessed and parcelled into 116 thickness and volume information for
anatomical structures (for the full list of features see
Table~\ref{tab:freesurfer_list}), according to the Desikan-Killiany atlas and
ASEG Freesurfer atlas~\citep{Desikan-2006An}. From these segmented regions, we
extracted the cortical thickness and volume to be the input data for our further
analysis.

\subsection{TPOT Automated Analysis} TPOT \citep{Olson-2016Automating,
Olson-2016Evaluation} uses genetic programming to search through different
operators (i.e. preprocessing approaches, machine learning models, and their
associated hyperparameters) to iteratively evolve the most suitable pipeline
with high accuracy. It does so by 1) generating a pool of random analysis models
sampled from a dictionary of preprocessing approaches and analysis models (See
Table~\ref{tab:algorithms} for a list of the models used); 2) evaluating these
models using 10-fold cross-validation, to identify the most accurate pipeline
with the lowest amount of operators; 3) breeding the top 20 selected pipelines
and applying local perturbations (e.g. mutation and cross-over); 4)
re-evaluating the pipeline in the next generation. This process is repeated for
a specified number of generations before settling on a final optimal pipeline
that has high accuracy and low complexity (i.e., lowest number of pipeline
operators). To make sure that the operators are combined in a flexible way, TPOT
uses a tree-based approach. That means that every pipeline is represented as a
tree where the nodes represented by the different operators. Every tree-based
pipeline starts with one or more copies of the dataset and every time the data
is passed through a node the resulting prediction is saved as a new feature. In
particular, TPOT uses a genetic programming algorithm as implemented in the
Python package DEAP (\cite{Fortin-2012DEAP:}; for a more detailed description of
the TPOT implementation see \cite{Olson-2016Evaluation}).

\subsubsection{Regression} \paragraph{TPOT hyperparameters exploration} We used
TPOT to find the 'best' pipelines to predict brain age, where the fitness of the
pipeline is defined by a low MAE between the predicted and the subject's
chronological age. To do this, we randomly selected 1546 subjects from the
dataset (TPOT training set), and we applied TPOT on them for 10 generations to
find the most fitted ML pipeline - the pipelines with the highest accuracy. The
optimal pipeline suggested by TPOT was then used to train an independent
(n=8761) dataset and its performance was evaluated using a 10-fold
cross-validation. The TPOT analysis and the evaluation of the model in an
independent training set was repeated 10 times. As a result, we obtained 100
performance scores for each configuration that were used to evaluate the impact
of manipulating a) the types of model preprocessing, b) number of models tested
on the first generation, and c) mutation and cross-over rate.

\paragraph{Comparison between TPOT and RVR} We also performed a 10 times
repetition with 10-fold cross-validation (as described above) to assess the
difference in performance between the 'best' pipelines yielded by TPOT and the
RVR, a standard model used in brain-age prediction \citep{Franke-2010Estimating,
Madan-2018Predicting, Kondo-2015An, Wang-2014Age}. In addition, to check if the
underlying age distribution would have an effect on the models yielded by TPOT,
we repeated the analysis using 784 subjects whose age was uniformly distributed
between 18-77 years old. In this case, we used n=117 subjects to train TPOT and
obtain the best pipeline. The remaining subjects (n=667) were used to train the
best pipeline using a 10-fold cross-validation. Similarly to the other analyses,
this evaluation process was also repeated 10 times resulting in 100 MAE values
for each condition.

While a Student's t-test is often used to check the difference in performance
between two models; Student's test assumes that samples are independent, an
assumption that is violated when performing a k-fold cross-validation. As part
of the k-fold cross-validation procedure, one subject will be used in the
training set k-1 times. Therefore, the estimated scores will be dependent on
each other, and there is a higher risk of type I error. For this reason, we used
a corrected version of the t-test that accounts for this dependency
\citep{Nadeau-2000Inference} when comparing the performance of TPOT and RVR and
the Friedman test when comparing different hyperparameters from TPOT
\citep{Demsar-2006Statistical}.

%%%%%%%%%%%%%%%%%%%%%%%%%%%%%%%%%%%%%%%%%%%%%%%%%%%%%%%%%%%%%%%%%%%%%%%%%%%%%%%%%%
% Results
%%%%%%%%%%%%%%%%%%%%%%%%%%%%%%%%%%%%%%%%%%%%%%%%%%%%%%%%%%%%%%%%%%%%%%%%%%%%%%%%%%%
\section{Results} 
\begin{figure}
    \includegraphics[width=\textwidth]{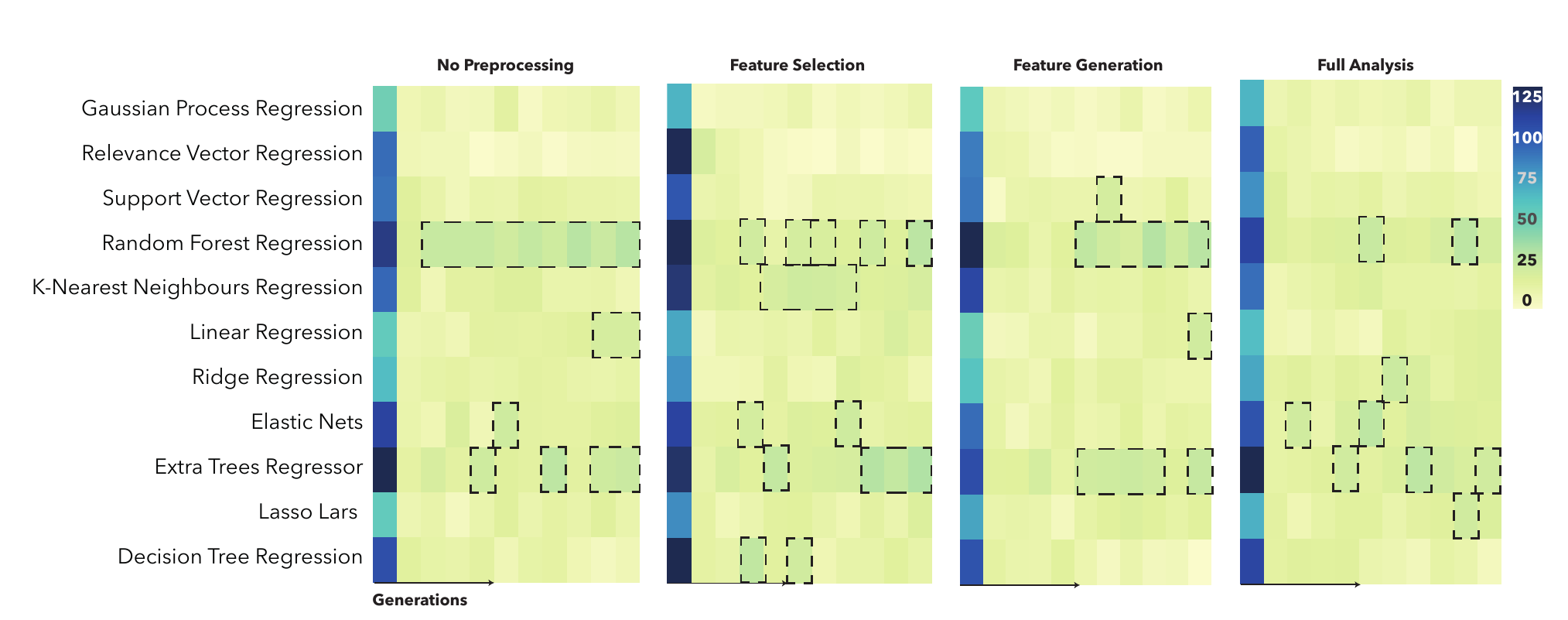}
    \caption{\textbf{Overview of the models count for each generation from one
    repetition for the different configurations experiments: A:} Models with a
    darker colour were more popular then models with lighter colour. Across the
    four experiments, Random Forest, K-Nearest Neighbours, Linear Regression and
    Extra Trees Regressors are the models with the highest count per generation.
    To make sure that all models were represented we increased the number of
    times the models were evaluated in the first generation.}
    \label{fig:complex_image2}
\end{figure}

We firstly investigated which models survived thought the
different generations. Figure~\ref{fig:complex_image2} shows the counts of the
different models in one of the repetitions. Random Forests and Extra Trees
Regressors are the most popular models followed by Elastic Nets. Decision Trees
and K-Nearest Neighbours also have a high popularity for the feature selection
configuration.

\subsection{TPOT parameter exploration}

\begin{figure}
    \includegraphics[width=\textwidth]{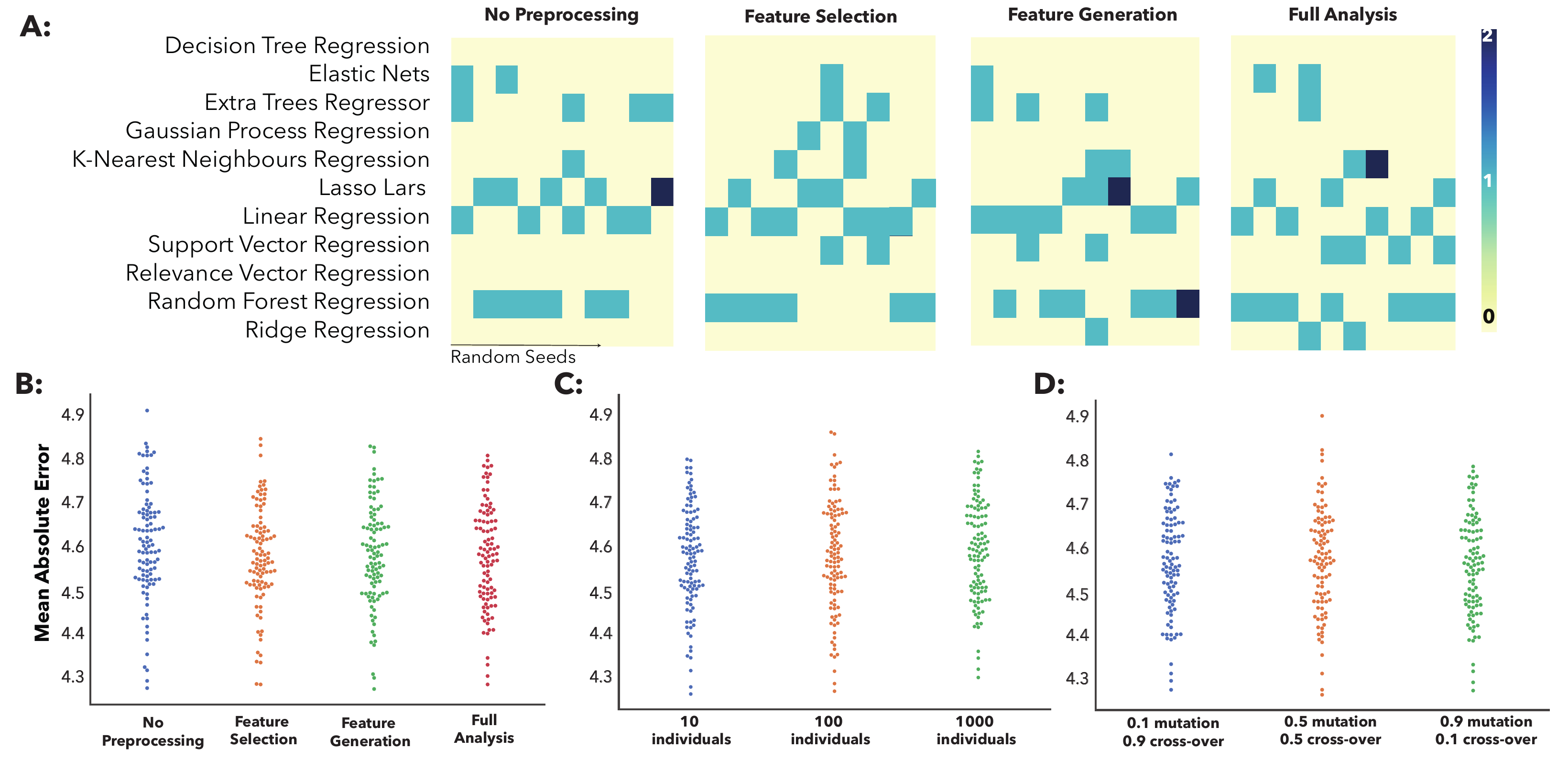}
    \caption{\textbf{Overview of the ensembles for the different analysis
    configurations at each repetition and their performance: A:} Schematic
    overview of the models composing the 'best' ensembles yielded by TPOT at
    each repetition. A darker colour represents models with higher counts.
    Random Forest Regression, Extra Trees Regressors, Lasso Lars and Linear
    Regression, were the most frequently represented. Despite the different
    models combinations among the different preprocessing analysis
    (\textbf{B:}), initial population size (\textbf{C:}), and
    mutation/cross-over rate (\textbf{D:}), there was no difference in the
    yielded performance.}
    \label{fig:boostrapped_fig}
\end{figure}

We then explored if the changes in the TPOT configuration are associated with a
different performance (Figure~\ref{fig:boostrapped_fig}B). We observed that
independent of the preprocessing we chose the performance varied between 4.3 and
4.9 years. In addition to that, for every repetition TPOT found a different
pipeline which was considered to be most accurate
(Figure~\ref{fig:boostrapped_fig}A). Similarly, we analysed the change in
performance when varying the initial population of pipelines
(Figure~\ref{fig:boostrapped_fig}C). If a model was not selected on the initial
population it will never be present in future generations, therefore we expected
that a larger initial population would lead to a more diverse pool and therefore
be associated with higher performances. We also explored the effect of mutation
and crossover rate on the performance of the derived pipelines. For a
combination of high (0.9), low (0.1), mid-ranges (0.5) mutation and cross-over
rates. (Figure~\ref{fig:boostrapped_fig}D). For all configurations, the
performance of the best models yielded by TPOT oscillated between 4.3 and 4.9
years. These suggest that there is not one single model that best describes the
dataset but a combination of many models leads to a higher performance and
independent of the of the underlying data structure TPOT was able to a pipeline
that yielded high performance.

\subsection{Comparison between TPOT and RVR} 
\begin{figure}
    \includegraphics[width=\textwidth]{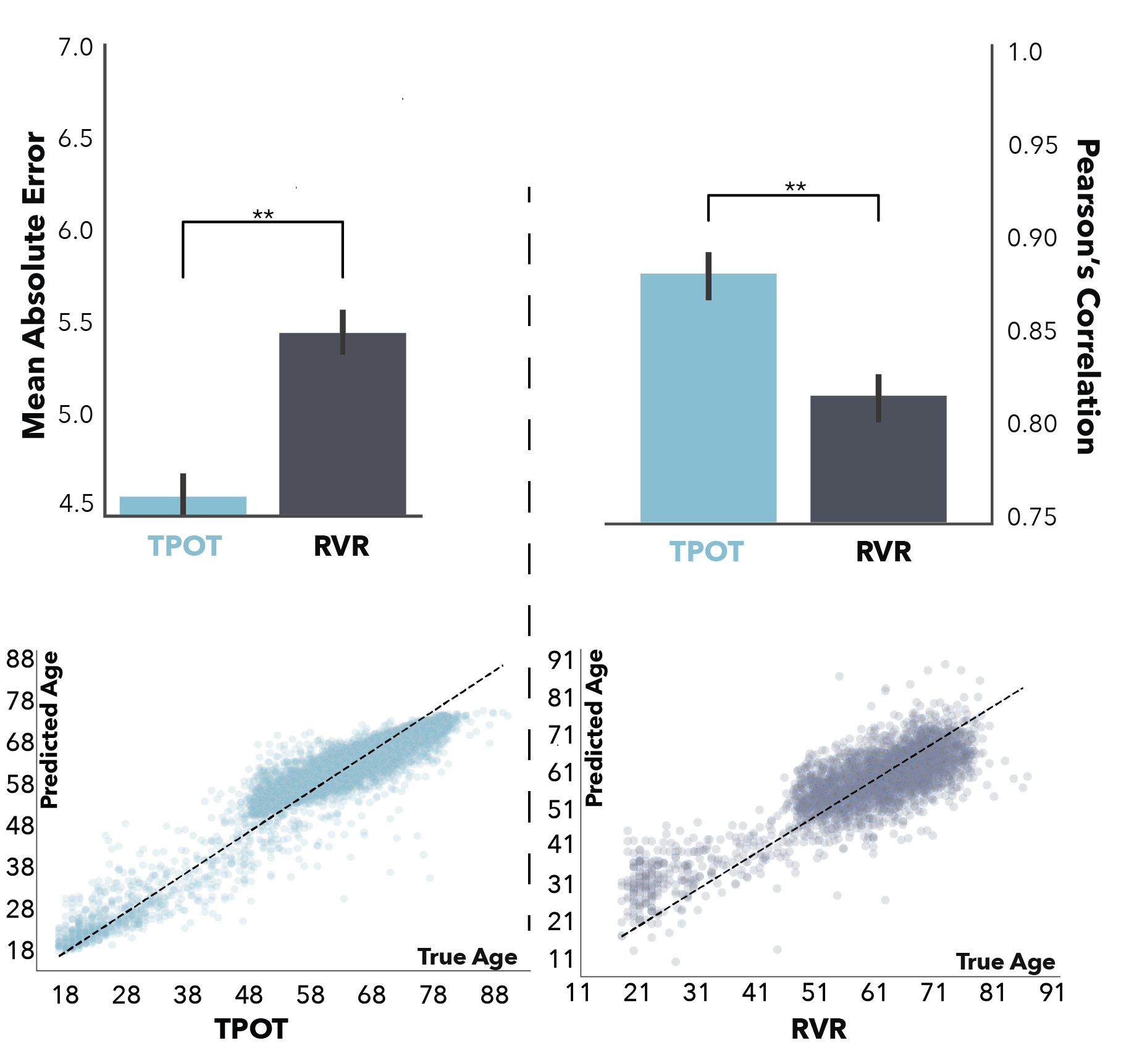}
    \caption{\textbf{Comparison of model's performance between TPOT and RVR:} We
    compared the MAE (top panel left) and Pearson's Correlation (top panel
    right) between True and Predicted age of the optimised model suggested by
    TPOT with and RVR on the test set. The lower panels show the Predicted vs
    the True age for one of the optimal pipelines suggested by TPOT (left) and
    RVR (right). Note that although both models use the same subject's to make
    prediction, the scales of the TPOT and RVR predictions are different, the
    RVR model predicts young subject's to be younger and old as older.
Asterisks show differences that are statistically significant at $p < 0.01$
(t-test corrected);  Error bars indicate $\pm 1SD$.}
    \label{fig:complex_image1}
\end{figure}

To assess the efficacy of the TPOT
approach applied to neuroimaging data, we compared the performance of the TPOT's
pipelines using the Full Analysis configuration with Relevance Vector
Regression. When using the entire dataset TPOT had a lower MAE and higher
Pearson's Correlation between true and predicted age
(Figure~\ref{fig:complex_image1}). However, when we applied TPOT to a uniformly
distributed dataset there was no significant difference between the models
yielded by TPOT and RVR (Table~\ref{tab:TPOTvsRVR}). Nevertheless, the models
suggested by TPOT using both datasets with the different age distribution were
similar (Figure~\ref{fig_sub:uniform}).

\begin{table}[]
\centering
\begin{tabular}{@{}clcclcc@{}}
\toprule
 & \multicolumn{1}{c}{MAE} & p-value & t & \multicolumn{1}{c}{Pearson's Correlation} & p-value & t \\ \midrule
TPOT &$4.612 \pm .124$ & \multirow{2}{*}{ \boldmath{$<.01$}} & \multirow{2}{*}{-6.441} & $.874 \pm .012$ & \multirow{2}{*}{ \boldmath{$<.01$}} & \multirow{2}{*}{3.745} \\
RVR & $5.474 \pm 0.140$ &  &  & $.813 \pm .0102$ &  &  \\
\midrule
\begin{tabular}[c]{@{}c@{}}TPOT\\ (uniform distribution)\end{tabular} & $5.594 \pm .0706$ & \multirow{2}{*}{$>.5$} & \multirow{2}{*}{-.616} & $.917 \pm .027$ & \multirow{2}{*}{$>.5$} & \multirow{2}{*}{.007} \\
\begin{tabular}[c]{@{}c@{}}RVR\\ (uniform distribution)\end{tabular} & $5.975 \pm .525$ &  &  & $.919 \pm .013$ &  &  \\
\bottomrule
\end{tabular}%
\caption{\textbf{Comparison between TPOT and RVR:} While TPOT has a significant higher accuracy and Pearson's Correlation when using the original data distribution, when using the uniformly distributed dataset both models had a similar performance. (The values represent $\pm SD$).}
\label{tab:TPOTvsRVR}
\end{table}

%%%%%%%%%%%%%%%%%%%%%%%%%%%%%%%%%%%%%%%%%%%%%%%%%%%%%%%%%%%%%%%%%%%%%%%%%%%%%%%%%%%%%%%%%%%%%%%%%%%

%%%%%%%%%%%%%%%%%%%%%%%%%%%%%%%%%%%%%%%%%%%%%%%%%%%%%%%%%%%%%%%%%%%%%
\section{Discussion}

The successful choice of an ML pipeline to predict variables of interest (such
as age) from neuroimaging data is driven by the statistical characteristics and
distribution of the dataset under analysis. In most cases, the choice of machine
learning model applied in multivariate analysis of neuroimaging data is rather
arbitrary - based on prior models that 'have worked', or by selecting whichever
model is most novel in the eyes of the analysis community. To explore an
alternative approach to model selection for a relatively simple problem, in this
work, we investigated the application of an automated analysis technique: TPOT.
The TPOT approach is a data-driven methodology which is agnostic to statistical
model \emph{and} prepossessing of the dataset - aiming to find the best pipeline
available to fit the statistical properties of the underlying dataset, whilst
simultaneously controlling for overfit and reliability. We showed that: (1) the
performance of the models suggested by TPOT is highly dependent on the specified
model pool (i.e. algorithms and hyperparameters) that TPOT has available to use.
However, feature selection, feature generation, initial population size the
mutation rate and cross-values rate do not have a substantial effect on the
TPOT's performance. (2) There is not one single machine learning algorithm that
performs the best, but good performance is achieved by a combination of models.
(3) The pipelines suggested by TPOT performed significantly better than commonly
used methods when performing a brain age regression from brain MRI scans.

Commonly used algorithms to predict brain age include a combination of linear
and non-linear ML algorithms such as: Multiple linear regression
\citep{Valizadeh-2017Age}, Gaussian Process Regressors
\citep{Cole-2015Prediction, Becker-2018Gaussian}, K-nearest neighbours
\citep{Valizadeh-2017Age}, Relevance Vector Regression \citep{Wang-2014Age,
Valizadeh-2017Age,  Franke-2010Estimating}, Random Forests
\citep{Valizadeh-2017Age} and Neural Networks \citep{Cole-2017Predictingb,
Valizadeh-2017Age}. In this study, we used an autoML approach that searched for
the most accurate pipeline over a pool of the commonly used algorithms and
compared its performance to RVR. We observed that the variance in the predicted
accuracy is very low on the test dataset for the pipelines suggested by TPOT but
also for the RVR model. This suggests that the models are not fitting to noise
but are finding interesting patterns in the data. Nevertheless, it is
interesting to note that for every analysis's repetition, a different pipeline
was yielded by TPOT which had the lowest MAE (i.e. 'best' pipeline;
Figure~\ref{fig:boostrapped_fig}). This is likely because there exists no single
model that always performs better for this type of regression problem.
Similarly, when analysing age prediction using voxel-wise data
\cite{Varikuti-2018Evaluation} previous work showed that the pattern of
'important' voxels is different across different training sets. Given the
strength of the association between brain structure and age, and high levels of
correlation between different brain regions, it seems that multiple different
approaches can achieve high levels of prediction accuracy. This cautions against
the over-interpretation of specific sets of model weights or coefficients as
being those specifically important for brain ageing, as it seems that a
different weighting on the brain could reach similar levels of performance.
Inference on which brain regions are most associated with ageing is better
conducted using a longitudinal within-subjects study design, rather than a
multivariate predictive model such as those used in TPOT. Our results also
highlight that all models yielded a similar MAE and were composed by a
combination of linear and non-linear models (Random Forest Regression, Extra
Tree Regression, K-Nearest Neighbours and Ridge or Lasso Regression;
Figure~\ref{fig:boostrapped_fig} and Figure~\ref{fig:complex_image2}). In
accordance with our results, \cite{Valizadeh-2017Age} also reported similar
brain-age prediction accuracy when comparing Random Forest and multiple linear
regression. One of the main advantages of Random Forests is that it can deal
with correlated predictors, while in a linear regression correlated predictors
might bias the results. Therefore, by combining both algorithms in an ensemble,
TPOT combines the strengths of both algorithms. Random forests have also been
used by \cite{Liem-2017Predicting} to combine multi-modal brain imaging data and
generate brain-age prediction. In particular, \cite{Liem-2017Predicting} used a
Linear Support Vector Regression to predict age and stacked these models with
Random Forests. This combined approach was able to improve brain-age prediction.
Our interpretation of these observations is that the use of Random Forests and
the hyperparameters found by TPOT 'better fit' the non-trival non-linearities
present in the dataset, transforming them within a n-dimensional manifold which
can then be fed trivially into a linear classifier. A similar observation has
been described by \cite{Aycheh-2018Biological}, where a combination of Sparse
Group Lasso and Gaussian process regression was used to predict brain age. On
the other hand, whilst stable, and able to generalise, this non-linear
transformation and combinations of different models into a pipeline makes
interpretation of important features within the dataset impossible.

We also noted that when using a subsample of the dataset that has a uniform
distribution, similar models were used by TPOT to build ensembles, nevertheless
the difference in performance between TPOT and RVR was not significant
(Table~\ref{tab:TPOTvsRVR}). We hypothesise that by using a uniform distribution
we make the problem of age regression easier and therefore obtained similar
performance between the TPOT and RVR approach, or that the reduced sample used
to pre-train TPOT was not sufficient to obtain an accurate fit. It would be
interesting for future research to explore these hypotheses further.

In the context of other literature, it is important to note that more accurate
brain-age prediction models \citep{Cole-2017Predictingb}, do exist. As shown by
\cite{Cole-2017Predictingb}, Convolutional Neural Networks can predict brain age
with a MAE of 4.16 years using a similar age range (18-90 years, mean
age=36.95). As developing neural networks requires in-depth knowledge of
architecture engineering, it would be interesting to use autoML approaches to
explore and select the most appropriate network architecture. However, the
approach in the present study does not make any assumptions about the underlying
statistics of the dataset and does not require any fine-tuning of the model of
choice but still achieve state-of-the-art accuracy. When comparing the accuracy
of different studies, it is important to take into account the age range of the
analysed sample, as age prediction in a small range has less variability than in
a large range. In fact,  using a sample with subjects aged 45-91
\cite{Aycheh-2018Biological} obtained a MAE of 4.02 years. While
\cite{Valizadeh-2017Age} had a similar age range as that described in our
project, they do not report the MAE for the entire sample and use instead 3 age
groups (8-18 years, 18-65 and 65-96 years) to test the accuracy of different
models. In general, \cite{Valizadeh-2017Age} reported lower accuracy for the
older group with MAE ranging between 4.90 and 14.23 years, when using only the
thickness information. On the other hand, \cite{Liem-2017Predicting} using only
the cortical thickness reported a MAE of 5.95 years (analysed age range 18 - 89
years, mean = 58.68).

In the specific case of Deep-Neural Network approaches to the brain age problem,
whilst improvements can be made on the accuracy of the model, often this is at
the cost of reliability.  As TPOT can accommodate a wider set of models, it
would be interesting to include Neural Networks on the model pool and compare
its performance against the range of selected models or to use other autoML
toolboxes like autokeras \citep{Jin-2019Auto-keras:} or Efficient Neural
Architecture Search via Parameter Sharing \citep{Pham-2018Efficient}. This
automated approach will allow an extensive search of models and parameters and
might also shed light into the question if deep learning is beneficial
neuroimaging analysis. Recently, \cite{Schulz-2019Deep} showed that linear,
kernels and deep learning models show very similar performance in brain-imaging
datasets. Combining the potential power of deep-learning with a model-agnostic
technique such as employed by TPOT, offers an potentially interesting route for
further research.

\section{Conclusion}
Overall, our results show that the TPOT approach can be used as a data-driven
approach to find ML models that accurately predict brain age. The models yielded
by TPOT were able to generalise to unseen dataset and had a significantly better
performance then RVR. This suggests that the autoML approach is able to adapt
efficiently to the statistical distribution of the data. Although more accurate
brain-age prediction models have been reported \citep{ Cole-2017Predictingb},
the approach in the present study uses a wide age range (18-89 years old), uses
only cortical anatomical measures, but most of all, it does not make any
assumptions about the underlying statistics of the dataset and does not require
any fine-tuning of the model of choice. By extensively testing different models
and its hyperparameters, TPOT will suggest the optimal model for the training
dataset. This approach removes possible introduced bias out of the loop and
allows decisions about the model to be made in an automated, data-driven and
reliable way.

%%%%%%%%%%%%%%%%%%%%%%%%%%%%%%%%%%%%%%%%%%%%%%%%%%%%%%%%%%%%%%%%%%%%%
\section{Acknowledgements}
We thank Sebastian Popescu for his help in carrying out the Freesurfer analysis
on the BANC dataset, and Pedro F. da Costa for enlightening discussions and
feedback on the analysis.

JD is funded by the King’s College London \& Imperial College London EPSRC
Centre for Doctoral Training in Medical Imaging (EP/L015226/1). WHLP was
supported by Wellcome Trust (208519/Z/17/Z). FET is funded by the PET
Methodology Program Grant (Ref G1100809/1) and the project grant “Development of
quantitative CNS PET imaging probes for the glutamate and GABA systems” from the
Medical Research Council UK (MR/K022733/1). SRC was supported by the Medical
Research Council (MR/M013111/1 and MR/R024065/1), the Age UK-funded Disconnected
Mind project (http://www.disconnectedmind.ed.ac.uk), and by a National
Institutes of Health (NIH) research grant (R01AG054628). PJH is supported by a
Sir Henry Wellcome Postdoctoral Fellowship from the Wellcome Trust
(WT/106092/Z/14/Z)

\paragraph{Author contribution}: JD, WHLP, FT, JHC, RL, PJH designed the study.
MAE, SRC, HCW, AMM, JHC preprocessed the dataset. JD performed the experiments.
JD, WHLP and PJH analysed the data. JD, PJH, WHLP, FT, JHC, SRC, HCW wrote and
edited the manuscript.

%%%%%%%%%%%%%%%%%%%%%%%%%%%%%%%%%%%%%%%%%%%%%%%%%%%%%%%%%%%%%%%%%%%%%
% Bibliography
%%%%%%%%%%%%%%%%%%%%%%%%%%%%%%%%%%%%%%%%%%%%%%%%%%%%%%%%%%%%%%%%%%%%%
\bibliography{bib/tpot_manuscript_new_version.bib}

% Appendix
\newpage
\section{Supplementary Info}

% Restart Table and Figure counts and name them with S
\setcounter{table}{0}
\renewcommand{\thetable}{S\arabic{table}}
\setcounter{figure}{0}
\renewcommand{\thefigure}{S\arabic{figure}}
%%%%%%%%%%%%%%%%%%%%%%%%%%%%%%%%%%%%%%%%%%%%%%%%%%%%%%%%%%%%

\begin{table}[H]
\begin{tabularx}{\linewidth}{ll}
%\centering
    \toprule
    Algorithms    &  Sklearn Implementation \\
    \hline
    \multicolumn{1}{l}{\emph{Feature Selection}}\\
    \hline
    Principle Component Analysis &  PCA \\
    Fast algorithm for Independent Component Analysis & FastICA \\
    Select the p-values corresponding to Family-wise error rate & SelectFwe \\
    Select features according to a percentile of the highest scores & SelectPercentile \\
    Remove low-variance Features & VarianceThreshold\\
    \hline
    \multicolumn{1}{l}{\emph{Feature Generation}}\\
    \hline
    Agglomerate features & FeatureAgglomeration \\
    \hline
    \multicolumn{1}{l}{\emph{Regression}}\\
    \hline
    Elastic Net model with iterative fitting along a regularisation path &   ElasticNetCV \\
    Randomised Decision Trees on sub-samples of the dataset  &   ExtraTreesRegressor \\
    k-Nearest Neighbours Regression          & KNeighborsRegressor \\
    Cross-validated Lasso using the LARS algorithm        &   LassoLarsCV \\
    Linear Support Vector Regression & LinearSVR \\
    Linear Least squares with l2 regularisation & Ridge \\
    Random Forest Regressor  & RandomForrestRegressor \\
    Ordinary Least Squares Linear Regression & LinearRegression \\
    Decision Tree Regressor & DecisionTreeRegressor \\
    Gaussian process regression  & GaussianProcessRegressors \\
    Relevance Vector Regression &  RVR \\
    \bottomrule
    \caption{List of used Feature Selection, Feature Generation and Regression Algorithms}
    \label{tab:algorithms}
 \end{tabularx}
 \end{table}

%%%%%%%%%%%%%%%%%%%%%%%%%%%%%%%%%%%%%%%%%%%%%%%%%%%%%%%%%%%%
\begin{sidewaysfigure*}
\thispagestyle{empty}
\begin{adjustwidth}{-5cm}{-5cm}
\includegraphics[page=1]{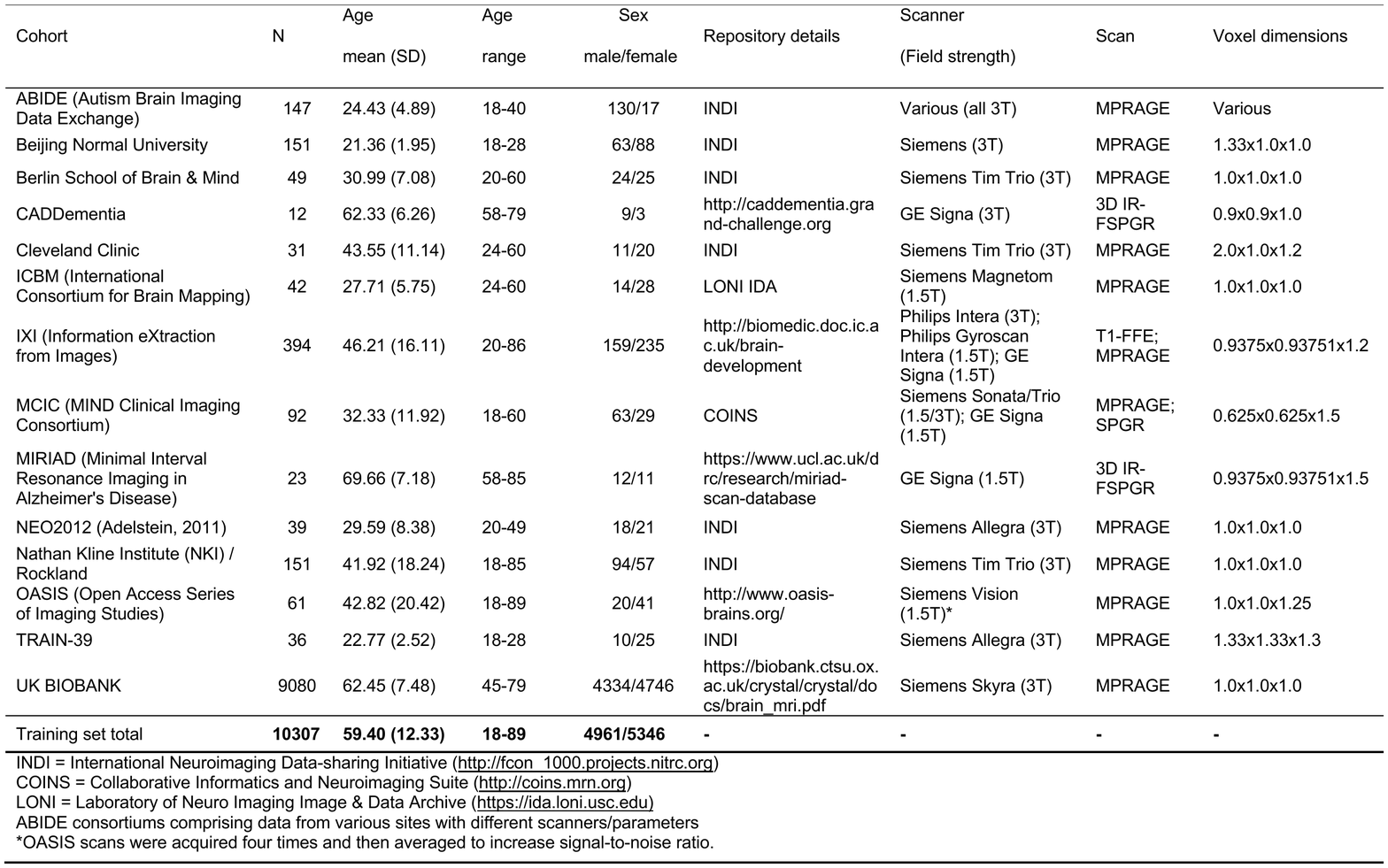}
\end{adjustwidth}
\vspace{-3cm}
\captionof{table}{Overview of the demographics and imaging parameters for the different datasets.}
\label{table:demography}
%\includepdf[landscape=true]{previous_versions/dataset_info}\label{table-demographics}

\end{sidewaysfigure*}

%%%%%%%%%%%%%%%%%%%%%%%%%%%%%%%%%%%%%%%%%%%%%%%%%%%%%%%%%%%%

\newpage
\begin{longtable}{lll}
    \caption{List of used Freesurfer features} \\ \toprule
    lh bankssts thickness &
    lh caudalanteriorcingulate thickness &
    lh caudalmiddlefrontal thickness \\
    lh cuneus thickness &
    lh entorhinal thickness&
    lh fusiform thickness \\
    lh inferiorparietal thickness &
    lh inferiortemporal thickness&
    lh isthmuscingulate thickness\\
    lh lateraloccipital thickness&
    lh lateralorbitofrontal thickness &
    lh lingual thickness \\
    lh medialorbitofrontal thickness &
    lh middletemporal thickness &
    lh parahippocampal thickness \\
    lh paracentral thickness &
    lh parsopercularis thickness &
    lh parsorbitalis thickness\\
    lh parstriangularis thickness &
    lh pericalcarine thickness &
    lh postcentral thickness \\
    lh posteriorcingulate thickness &
    lh precentral thickness &
    lh precuneus thickness \\
    lh rostralanteriorcingulate thickness &
    lh rostralmiddlefrontal thickness &
    lh superiorfrontal thickness \\
    lh superiorparietal thickness &
    lh superiortemporal thickness &
    lh supramarginal thickness \\
    lh frontalpole thickness &
    lh temporalpole thickness &
    lh transversetemporal thickness \\
    lh insula thickness &
    lh MeanThickness thickness &
    rh bankssts thickness \\
    rh caudalanteriorcingulate thickness &
    rh caudalmiddlefrontal thickness &
    rh cuneus thickness \\
    rh entorhinal thickness &
    rh fusiform thickness &
    rh inferiorparietal thickness \\
    rh inferiortemporal thickness &
    rh isthmuscingulate thickness &
    rh lateraloccipital thickness \\
    rh lateralorbitofrontal thickness &
    rh lingual thickness &
    rh medialorbitofrontal thickness \\
    rh middletemporal thickness &
    rh parahippocampal thickness &
    rh paracentral thickness \\
    rh parsopercularis thickness &
    rh parsorbitalis thickness &
    rh parstriangularis thickness \\
    rh pericalcarine thickness &
    rh postcentral thickness &
    rh posteriorcingulate thickness \\
    rh precentral thickness &
    rh precuneus thickness &
    rh rostralanteriorcingulate thickness \\
    rh rostralmiddlefrontal thickness &
    rh superiorfrontal thickness &
    rh superiorparietal thickness \\
    rh superiortemporal thickness &
    rh supramarginal thickness &
    rh frontalpole thickness \\
    rh temporalpole thickness &
    rh transversetemporal thickness &
    rh insula thickness \\
    Left-Cerebellum-White-Matter &
    Left-Cerebellum-Cortex &
    rh MeanThickness thickness \\
    Left-Thalamus-Proper &
    Left-Caudate &
    Left-Putamen \\
    Left-Pallidum &
    3rd-Ventricle &
    4th-Ventricle \\
    Brain-Stem &
    Left-Hippocampus &
    Left-Amygdala \\
    CSF &
    Left-Accumbens-area &
    Left-VentralDC \\
    Left-vessel &
    Right-Cerebellum-White-Matter &
    Right-Cerebellum-Cortex \\
    Right-Thalamus-Proper &
    Right-Caudate &
    Right-Putamen \\
    Right-Pallidum &
    Right-Hippocampus &
    Right-Amygdala \\
    Right-Accumbens-area &
    Right-VentralDC &
    Right-vessel \\
    CC Posterior &
    CC Mid Posterior &
    CC Central \\
    CC Mid Anterior &
    CC Anterior &
    rhCortexVol \\
    CortexVol &
    lhCerebralWhiteMatterVol &
    rhCerebralWhiteMatterVol \\
    CerebralWhiteMatterVol &
    SubCortGrayVol &
    TotalGrayVol \\
    SupraTentorialVol &
    SupraTentorialVolNotVent &
    SupraTentorialVolNotVentVox \\
    MaskVol &
    BrainSegVol-to-eTIV &
    MaskVol-to-eTIV \\
    EstimatedTotalIntraCranialVol &
    &
     \\
    \bottomrule
    \label{tab:freesurfer_list}
\end{longtable}

%%%%%%%%%%%%%%%%%%%%%%%%%%%%%%%%%%%%%%%%%%%%%%%%%%%%%%%%%%%%
\newpage
\begin{figure}
    \includegraphics[width=\textwidth]{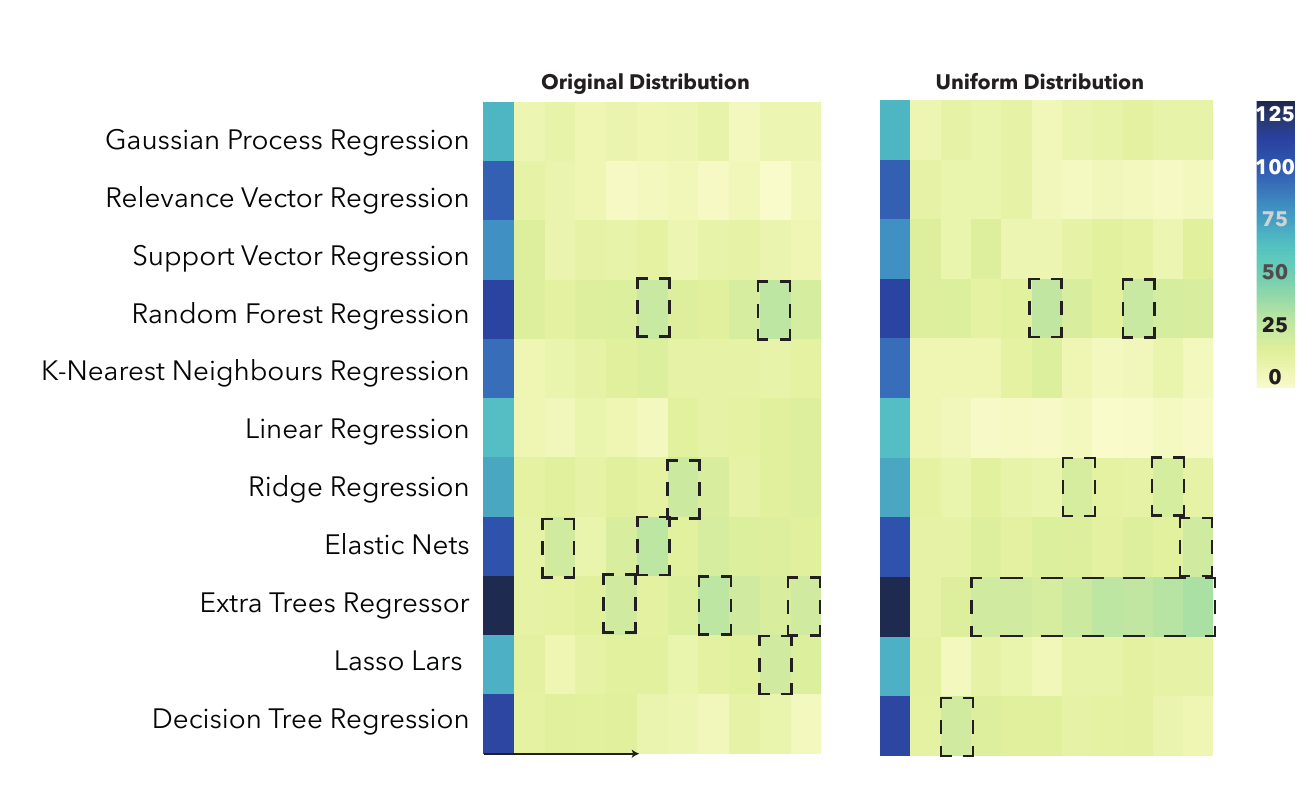}
    \caption{Model counts for the different distributions for one single repetition over the different generations. While \textbf{Left} shows the model count for the normal data distribution, \textbf{Right} illustrates the model counts for the uniform distribution. Using both distributions, the most common models explored by TPOT are Random Forrest Regression, Extra Tree Regressors and Elastic Nets.}
    \label{fig_sub:uniform}
\end{figure}

\end{document}